\documentclass[journal]{IEEEtran}

\usepackage{cite}
\usepackage{amsmath}
\usepackage{textcomp}
\usepackage{wasysym}

\usepackage{graphicx}

\usepackage{tikz}

\usepackage{siunitx}

\begin{document}

\title{Experimental Synthetic Aperture Radar with Dynamic Metasurfaces}

\author{Timothy~Sleasman*,~\IEEEmembership{Student Member,~IEEE, }%
	Michael~Boyarsky,
	Laura~Pulido-Mancera,~\IEEEmembership{Student Member,~IEEE, }%
	Thomas~Fromenteze,
	Mohammadreza~F.~Imani,~\IEEEmembership{Member,~IEEE, }%
	Matthew~Reynolds,~\IEEEmembership{Senior Member,~IEEE, }%
	and~David~R.~Smith%
\thanks{T. Sleasman, M. Boyarsky, L. Pulido-Mancera, T. Fromenteze, M. F. Imani, and D. R. Smith are with the Center for Metamaterials and Integrated Plasmonics, Duke University, Department of Electrical and Computer Engineering, Durham, NC, 27708 USA. M. Reynolds is with the University of Washington, Department of Electrical Engineering, Seattle, WA 98195, USA.}%
\thanks{Corresponding author: sleasmant@gmail.com}}

\markboth{IEEE TAP,~Vol.~PP, No.~PP, July~2016}%
{IEEE TAP,~Vol.~PP, No.~PP, July~2016}

\maketitle

\begin{abstract}
We investigate the use of a dynamic metasurface as the transmitting antenna for a synthetic aperture radar (SAR) imaging system. The dynamic metasurface consists of a one-dimensional microstrip waveguide with complementary electric resonator (cELC) elements patterned into the upper conductor. Integrated into each of the cELCs are two diodes that can be used to shift each cELC resonance out of band with an applied voltage. The aperture is designed to operate at K band frequencies (17.5 to 20.3 GHz), with a bandwidth of 2.8 GHz. We experimentally demonstrate imaging with a fabricated metasurface aperture using existing SAR modalities, showing image quality comparable to traditional antennas. The agility of this aperture allows it to operate in spotlight and stripmap SAR modes, as well as in a third modality inspired by computational imaging strategies. We describe its operation in detail, demonstrate high-quality imaging in both 2D and 3D, and examine various trade-offs governing the integration of dynamic metasurfaces in future SAR imaging platforms.
\end{abstract}

\begin{IEEEkeywords}
Synthetic aperture radar, Microwave imaging, Radar imaging, Beam steering, Reconfigurable antennas, Waveguide antennas
\end{IEEEkeywords}

\section{Introduction}
\label{sec:sec1}
\IEEEPARstart{S}{ynthetic} aperture radar (SAR) is an established technology that has been applied in aircraft- and satellite-based imaging since the mid-twentieth century \cite{curlander1991SAR,soumekh1999SAR,brown1969SARintroduction}. In its earliest realizations, a simple antenna with a directive beam was affixed to a moving platform that traveled along a large distance \cite{soumekh1999SAR}. This configuration allows for the synthesis of an effective aperture that is much larger than what would be practical using stationary antennas. High-resolution images in SAR systems are thus achieved by using frequency bandwidth (to sense objects along the range coordinate) coupled with a spatially-large aperture (to sense objects along the cross-range coordinate). Moreover, given the inherent Fourier transform relationship between data sampled over a synthetic aperture and the scene reflectivity, fast and efficient image reconstruction techniques are available, making SAR a powerful and preferred imaging platform \cite{Pulido-Mancera2016c,soumekh1994fourierarray,yarovoySparseSAR,zhuge2012three}. Its many applications include security screening \cite{sheen2001threatimaging,alvarez2015_3dCS,gonzalez2014sparse,ahmed2011Imaging,ahmed2012Imaging}, space and earth observation \cite{sarabandi1992EarthImage,treuhaft1996vegetation}, through-wall imaging \cite{dehmollaian2008throughwall,huang2010throughwall}, among others.

\begin{figure} 
	\centering
	\includegraphics[width=3.0in]{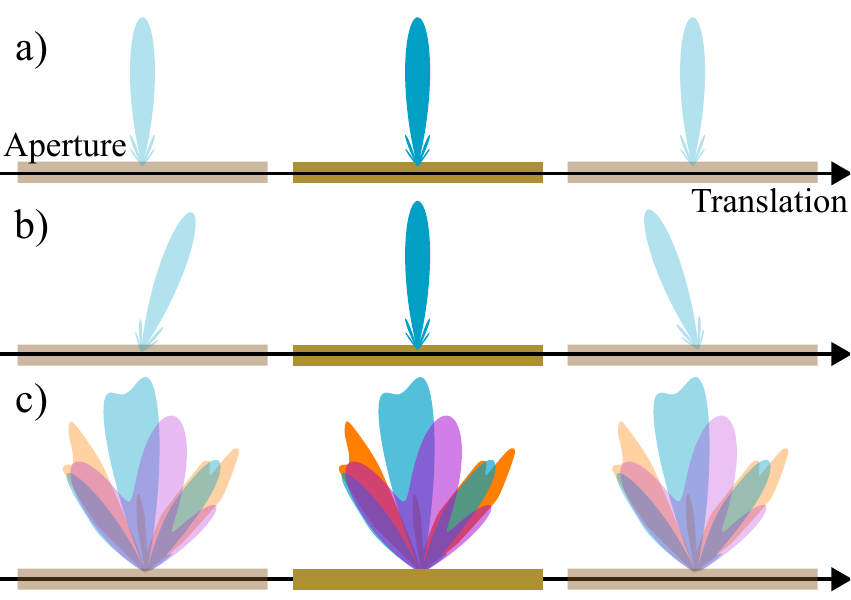}
	\caption{\label{fig:f1}a) Stripmap and b) spotlight SAR imaging, with the arrow indicating the path of motion. Alternatively, c) diverse pattern stripmap SAR can be executed with complex patterns.}
\end{figure}

A variety of antennas with different radiative characteristics have been used for SAR, with several common imaging modalities becoming highly developed. In the simplest scenario---depicted in Fig. \ref{fig:f1}a and commonly referred to as stripmap SAR---the direction of the antenna's beam is held constant and the region of interest is dictated by the path of travel \cite{soumekh1999SAR}. This method has a simple architecture but lacks the ability to actively direct the antenna at a specific location, an inadequacy which leads to a trade-off between resolution and signal-to-noise ratio (SNR) \cite{soumekh1999SAR}. Spotlight SAR, in which the antenna's beam is continuously adjusted to aim at a fixed point of interest, can increase resolution by synthesizing a larger effective aperture relative to the imaging domain \cite{jakowatz2012spotlight}. This mode, however, requires a more complex architecture to steer the antenna and results in a limited region of interest.  The spotlight SAR modality is depicted in Fig. \ref{fig:f1}b.

Depending on the specific SAR application, a multitude of hardware platforms have been demonstrated, including phased arrays \cite{hansen2009phasedarray} and gimbal-actuated parabolic reflector dishes. Phased arrays and electronically scanned antennas (ESAs) \cite{johnson2015sidelobeCancel,patel2014ESA,patel2015ESA,lim2004CRLHsteer} possess near-total control over radiated fields but often have a costly, heavy, or complex architecture that consumes significant power. Reflector dishes offer near-perfect radiation efficiency and can generate highly-directive beams; however, they suffer from a large physical footprint, offer minimal control over the radiated fields, and rely on mechanical rotation to steer the beam. While some applications can adequately exploit the benefits of these specific antenna architectures, flexibility in field patterns and a reduced physical footprint stand to provide substantial advantages for many imaging environments.

To enable tailoring of radiation patterns from a low-cost and planar platform, waveguide-fed metasurface antennas may be considered \cite{hunt2013science}. These antennas consist of a waveguide (e.g. a microstrip or parallel plate waveguide) which excites a multitude of subwavelength, resonant radiators. Each resonator couples a portion of the guided mode's energy to free space \cite{landy2013WGhomogenization}. The overall radiation pattern generated by this aperture is thus the superposition of the radiation from each single radiator. By integrating a tuning mechanism into each independent resonator, further control over the radiated waveforms can be achieved \cite{sleasman2015DyAp1}. The flexibility offered by dynamic metasurfaces may be used to steer directive beams for enhanced signal strength \cite{lim2004CRLHsteer}, create nulls in the pattern to avoid jamming \cite{jammingSAR,jammingSAR2}, probe a large region of interest with a wide beam, or even interrogate multiple positions at once with a collection of beams \cite{mimosar}.

Similar to leaky wave antennas \cite{oliner1993leaky}, metasurface apertures naturally produce radiation patterns that vary as a function of frequency. In a relatively recent application, a passive, frequency-diverse metasurface aperture has been used as a novel imaging device in the context of security screening and threat detection \cite{hunt2014metaimager}. In this application, an electrically-large metasurface aperture---consisting of a parallel plate waveguide patterned with arrays of cELCs whose resonance frequencies are chosen randomly over the bandwidth of operation---produces a sequence of distinct and complex radiation patterns as a function of driving frequency. From the measured return signal of these modes, high-fidelity images can be reconstructed \cite{hunt2014metaimager}.

With the addition of dynamic tuning, the frequency dispersive nature of metasurface apertures can be mitigated. This can be applied, for example, to form directed beams or other patterns that are relatively frequency-invariant over a particular bandwidth. Dynamic metasurface apertures were first explored as a novel device used to form highly-directed beams for satellite communication and tracking \cite{johnson2015sidelobeCancel}.  More recently, a dynamic metasurface aperture was utilized in near-field computational imaging. In this case, uncorrelated patterns were employed to perform microwave imaging in a similar vein to the frequency-diverse approach pioneered in \cite{sleasman2015DyAp1,sleasmanMarathon}.

Dynamic metasurface aperture have proven their ability to generate a multitude of desired waveforms. As we will show, these waveforms can be applied in traditional SAR modalities, such as stripmap or spotlight SAR, both of which rely on specific beam patterns that are uniform across the operational bandwidth. Rather than strictly focusing on beam synthesis at a specific angle, we will also investigate other schemes that can utilize alternative waveforms---including uncorrelated, random patterns---for SAR imaging.

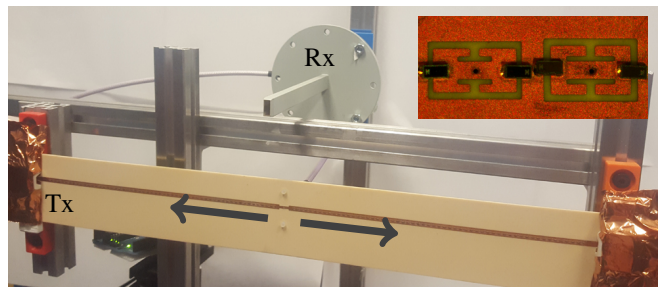
\begin{figure}
	\centering
	\begin{tikzpicture}
	\node[anchor=south west,inner sep=0] (image) at (0,0) {\includegraphics[width=3.4in]{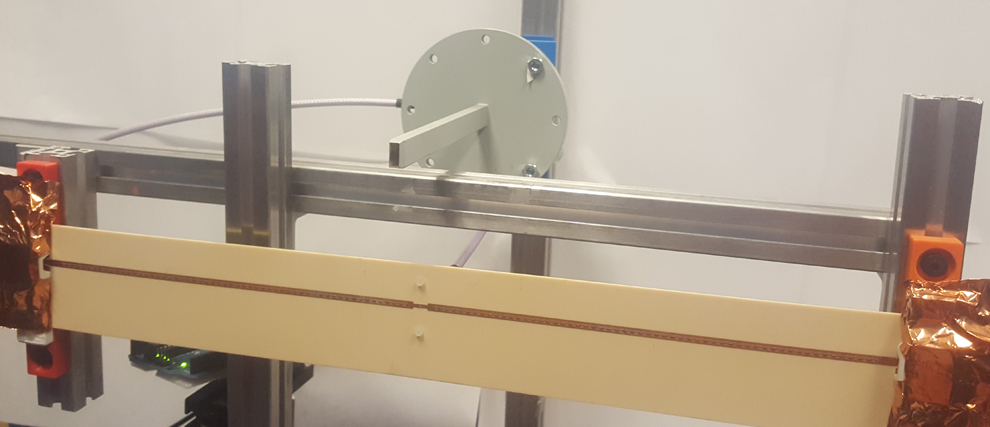}};
	\begin{scope}[x={(image.south east)},y={(image.north west)}]

	\node[anchor=south west,inner sep=0] (image) at (0.63,0.60) {\includegraphics[width=1.2in]{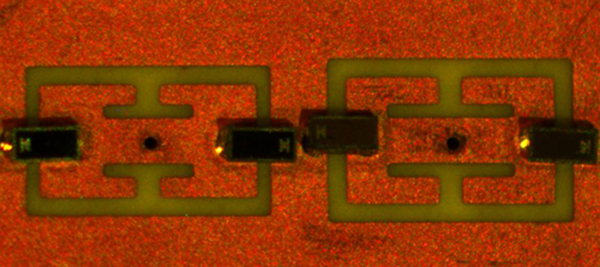}};
	
	\node at (0.48,0.82) {Rx};
	\node at (0.08,0.3) {Tx};
	\draw [line width=0.1cm,->,darkgray] (.45,.26-.03) -- (.6,.22-.03);
	\draw [line width=0.1cm,<-,darkgray] (.45-.2,.26+.025) -- (.6-.2,.22+.025);
	
	\end{scope}
	\end{tikzpicture}
	\caption{\label{fig:f2}The transmitting metamaterial aperture and the receiving open-ended waveguide. The inset shows a zoom in of the until cells and the tuning diodes. The arrows indicate the origin and the propagation direction of the guided wave.}
\end{figure}

In one specific non-standard SAR implementation, we consider physically translating the metasurface aperture along a path while forming random radiation patterns; we term this new SAR modality \emph{diverse pattern stripmap}. This scenario, depicted in Fig. \ref{fig:f1}c, is a fusion of computational imaging concepts \cite{brady2009CSholo,brady2009optical,donoho2006CS,watts2014THzCS,Caorsi2000compImag} and conventional SAR, leveraging the benefits of both. The large effective aperture is obtained from SAR and a multiplexing advantage is acquired from computational imaging \cite{fenimore1978codedaperture}. The main advantage of using these random patterns is that they can illuminate multiple points in the scene simultaneously, decoding the additional information in post processing. In this sense, the aperture probes many parts of the scene's spatial content simultaneously and investigates each location multiple times. We will qualify the pros and cons of this approach and contrast this method with the conventional SAR modalities.

In this manuscript, our goal is to expand the work of \cite{sleasmanMarathon} and investigate the utility of dynamic apertures in SAR applications. For this purpose, we take advantage of the dynamic metasurface presented in \cite{sleasmanMarathon}, which is shown in Fig. \ref{fig:f2}. While this device was originally designed to generate random patterns, we will see that its agile nature lends itself to be adapted for beam synthesis in spotlight and stripmap SAR imaging. Furthermore, it provides a platform to investigate how metasurface antennas in general can expand SAR capabilities and open new research avenues.

This paper begins with a review of the dynamic metasurface described in \cite{sleasmanMarathon} and a discussion of how it can generate various radiation patterns. Conventional stripmap and spotlight SAR modalities are then demonstrated with the device. Subsequently, diverse pattern stripmap is experimentally demonstrated, with random patterns used to multiplex the scene information as the aperture is physically translated over a linear path. Next, the three modalities are compared in studies that contrast the resolution and field of view. Finally, the aperture is physically translated in the perpendicular direction to perform volumetric imaging. Two- and three-dimensional reconstructions of point scatterers and complex objects are presented to explore the resolution and robustness of the various experimental techniques.

\section{Dynamic Metasurface Architecture}
\label{sec:sec2}

The dynamic aperture considered in this paper was designed and fabricated for use in a computational imaging framework. The design principles and specific fabrication considerations are detailed in \cite{sleasmanMarathon}; here we briefly review the aperture's operation and discuss its integration in the proposed SAR setting. In particular, this aperture was designed to generate uncorrelated radiation patterns that varied as a function of tuning state. By contrast, here we illustrate how this aperture can be tuned to generate steered, directive beams, for conventional SAR imaging, as well as random patterns, for the diverse pattern SAR modality introduced within this work.

The aperture, shown in Fig. \ref{fig:f2}, is a microstrip waveguide with cELC elements etched into its upper conductor \cite{hand2008cELC,sleasman2015element}. The cELC elements are chosen for the design since they behave as resonant, polarizable magnetic dipoles, and thus form efficient radiators in the otherwise electric surface \cite{landy2013WGhomogenization}. Moreover, each cELC can easily be made tunable by including two PIN diodes across the gaps in the straight segments, as shown in the inset to Fig. \ref{fig:f2}. The diodes act either as capacitors or resistive shorts depending on a binary DC voltage (\SI{0}{\volt} or \SI{5}{\volt}) applied via DC bias lines \cite{sleasman2015element}. The metasurface is designed such that when the PIN diodes are in the capacitive state the cELCs resonate, couple to the transverse magnetic field of the microstrip feed mode, and ultimately radiating energy into free space. When shorted, the resonance of the cELC elements is pushed out of band and the cELCs do not radiate, instead allowing the guided mode to propagate along the guide to subsequent elements in the aperture. These states are referred to as \emph{on} and \emph{off}, respectively. A given pattern of \emph{on}/\emph{off} elements will be referred to as a \emph{mask}.

Two separate cELC designs are etched into the waveguide with slightly different resonance characteristics. The resonance frequencies of the two elements are slightly offset, at \SIrange{17.8}{19.0}{\GHz}, as a means of extending the bandwidth and providing diversity in the aperture's radiated power spectrum. In this work, the aperture is fed at the center, resulting in a pair of counter-propagating feed waves within the microstrip. In total the aperture spans \SI{40}{\centi\meter} and contains 112 elements with spacing of \SI{3.33}{\milli\meter} ($\lambda_0/4$).

In many other aperture antenna platforms, such as phased arrays and leaky wave antennas, array factor calculations \cite{balanis2005antenna} and dispersion relations \cite{oliner1993leaky} inform the beam synthesis process. These methods provide weightings (i.e. phase and/or magnitude) for each of the radiating sites along the aperture which should be used to achieve a beam at a desired angle.

Since independent control over the phase and amplitude of each resonator is not available in metasurface antenna designs presented to date, metasurface apertures instead rely on the phase advance of the feed wave combined with the tuning state of the elements to provide beamforming capabilities. In this scenario, the waveguide mode can be thought of as a reference wave with the tuning states of the elements serving as a dynamically-reconfigurable computer-generated hologram \cite{goodmanFourierOptics}. In \cite{johnson2015sidelobeCancel} and \cite{johnson2014DDA_ApOp}, this type of holographic analysis was applied to generate steerable, directive beams using a dynamic metasurface. While this notion provides a more rigorous and sophisticated approach to understand the aperture's performance, we can also consider the aperture from an array factor perspective to develop some rudimentary intuition.

The weights for the sources implemented in the dynamic metasurface can be represented in the array factor calculation as

\begin{equation}
AF(\phi_0) = \sum_{m=1}^N A_m(\omega) e^{-j k_0 y_m sin(\phi_0)} e^{-j \beta y_m} 
\end{equation}

\noindent When this factor is multiplied by the individual element's radiation pattern \cite{balanis2005antenna,Pulido-Mancera2016}, the result is ideally a directive beam in the far field. In this model, $y_m$ is the position along the aperture, $k_0$ and $\beta$ are the wavenumbers of free space and the feed wave, $\omega$ is the frequency, $N$ is the number of elements, and $\phi_0$ is the desired beam angle. This equation can be solved for the elements' complex weighting factors, $A_m$, which can be enforced to achieve the desired pattern. Phased arrays directly control $A_m$ while our aperture has $A_m$ jointly determined by the feed waves and the tuning state of the elements. In our case, a binary square wave (i.e. with sequential elements are turned \emph{on} and \emph{off} with a given periodicity) for the $A_m$ values will enable many of the desired angles to be reached. The same result is found when considering the problem from a Fourier optics or holography perspective.

It should be emphasized that while this treatment gives a rough idea of the aperture's operation, it falls short in terms of delivering a predictive model. Strong inter-element coupling and fabrication inconsistencies within the aperture severely complicate many predictive analytic formulations. Further, the entire angular space cannot be covered with the array factor solution due to the limited control we possess over the elements. Thus, even though we will employ the square-wave masks that we know results in steered beams, we must also supplement these with further patterns.

In addition to square-wave masks, which each possess various periodicities, we have sampled a large variety of pseudo-random masks in an effort to generate additional steered beams. By sorting through many measurements of random masks, several radiation patterns matching the desired beam profiles can be found. This was completed through experimental near field scanning and is a cumbersome approach to the problem, but the one-time process ultimately results in satisfactory beams that fill many of the gaps in the frequency-angle coverage that otherwise result.

It should also be noted that since our specific aperture was designed for computational imaging, not beamsteering, the achievable beam characteristics (such as beamwidth and sidelobe levels) were limited. While this aperture is sufficient for initial experimental results, its design could be better tailored towards the goal of beam synthesis by applying techniques such as grayscale tuning, denser element placement, and tapered element design.

\begin{figure} 
	\centering
	\includegraphics[width=3.4in]{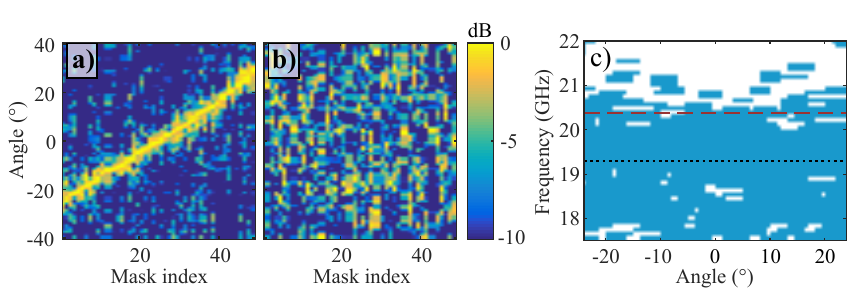}
	\caption{\label{fig:f3}a) The angle-frequency space coverage for steered beams. Directivity plots at \SI{19.3}{\giga\hertz} (the dotted line) are shown for a set of b) steered beams and c) random patterns. The useful bandwidth extends up to \SI{20.3}{\giga\hertz}, the dashed line in (a).}
\end{figure}

Despite the many obstacles that impede the beam synthesis process, the dexterity of the aperture, granted by the tunable metamaterial cells, redeems the structure. With the 112 elements all individually addressable, an enormous variety ($2^{112}$) of masks are possible and allow for the creation of the desired radiation patterns. As shown in Fig. \ref{fig:f3}a, the dynamic aperture masks can be selected such that a directive beam is steered over angles within \SI{\pm 23}{\degree}. The beams have a mean width of \SI{10}{\degree} and all have sidelobes of less than \SI{-4}{\dB}. An additional point of importance is the bandwidth for which this operation can be sustained. For spotlight SAR, the directive beam should ideally point at the desired angle for the entire operational bandwidth so that as the mechanical platform translates, it can always aim at the point of interest. As shown in Fig. \ref{fig:f3}c, a bandwidth of \SI{2.8}{\giga\hertz} (\SIrange{17.5}{20.3}{\GHz}) is achievable. A few angle-frequency pairs have not been realized, but this does not severely hinder the performance in this work and could be corrected with further exploration of additional masks.

Alternatively, the dispersive nature of the cells and their tunability can be exploited to remove the restriction of beam-like patterns. For example, a sampling of random patterns are shown in Fig. \ref{fig:f3}b; these patterns will be utilized in the diverse pattern stripmap SAR measurements introduced in section \ref{sec:sec1} and demonstrated in section \ref{sec:sec4}.

\section{Imaging Process}
\label{sec:sec3}

\begin{figure} 
	\centering
	\includegraphics[width=3.0in]{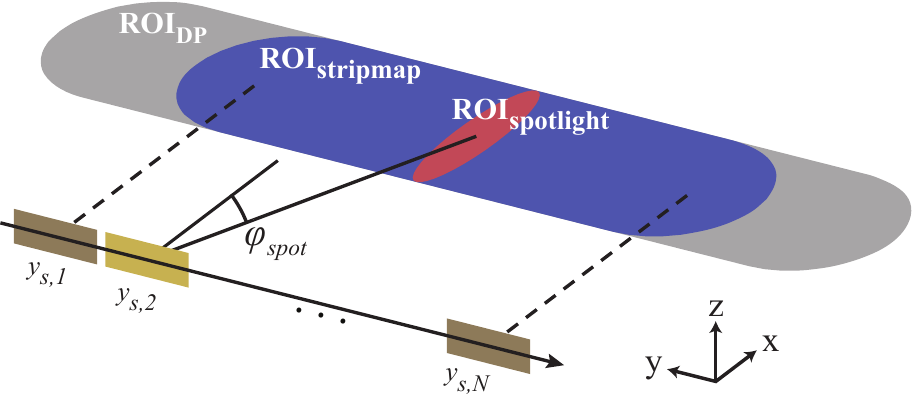}
	\caption{\label{fig:f4}The translating aperture (confined within the dashed lines) and the definition of the steered angle for spotlight imaging. Also shown are the inherent regions of interest for stripmap, spotlight, and diverse pattern stripmap.}
\end{figure}

Once the voltage tunings for different beam styles are selected, we can reproduce the desired patterns by invoking the corresponding masks. In our imaging configuration, the dynamic metasurface operates as a transmitter to illuminate a region of interest and the backscattered signal is collected by an open-ended waveguide receiver. Depending on the radiation patterns and the SAR modality, the region of interest will take on a different size, as shown in Fig. \ref{fig:f4}---this notion will become clear when the individual modalities are demonstrated. The transmit and receive antennas are held fixed and the scene translated in front of them, forming a quasi-monostatic, inverse SAR (iSAR) measurement \cite{soumekh1999SAR}---this is equivalent to the strategy shown in Fig. \ref{fig:f4} which provides a schematic with the key features. A Keysight E8653A vector network analyzer is used to measure the complex signal and the scene is translated on a Newmark D-slide 2D stage---both devices, as well as the voltages for the metamaterial cells, are controlled through MATLAB.

A key element in any imaging system is the description of the image processing, or forward model \cite{lipworth2013JOSA}. To properly process the backscattered signal, a model to relate the measurements with the radiated fields and the reflectivity of the scene needs to be developed. Various approaches have been pursued in the literature for this purpose, ranging from simple Born and Rayleigh approximations\cite{li2010validityBorn,lin1990invScatBorn} to non-linear models accounting for the materials properties of objects in the scene \cite{fear2002MWimaging,liu2004nonlinearRecon}. The choice of imaging model is usually determined by the intended application. In many SAR imaging systems, the aim is to retrieve a reflectivity map of the scene (qualitative imaging), which can be simply found with the Born approximation or Rayleigh formulas \cite{lin1990invScatBorn}. We adopt this framework and focus our efforts on the hardware aspects of the process. This method, based on computational imaging, is extremely versatile which enables its implementation throughout the various imaging styles explored in the work.

The formulation utilized here is heavily detailed in \cite{lipworth2013JOSA,lipworth2015virtualizer} and is employed for the aperture at hand in \cite{sleasmanMarathon}. In this framework, a standard matrix formulation concisely described the forward model as

\begin{equation}
\label{eq:eq2}
\overline{g} = \overline{\overline{H}} \, \overline{f} + \overline{n}
\end{equation}

\noindent where $\overline{f}$ is a vector describing the scene reflectivity, $\overline{g}$ is the measurement vector, and $\overline{n}$ is a noise term. The sensing matrix, $\overline{\overline{H}}$, is the cornerstone of the method and relates each generated field pattern to a single measurement \cite{lipworth2015virtualizer}. Writing out a single row of $\overline{\overline{H}}$ (and neglecting noise) we can see how the scene voxels and fields are related to a single measurement as

\begin{equation}
\label{eq:eq3}
g_i=\Sigma_jH_{ij}f_j=\Sigma_j(\overline{E}^T_i(\overline{r}_j) \cdot \overline{E}^R_i(\overline{r}_j))f_j
\end{equation}

\noindent Here, we have explicitly expanded the elements of the $\overline{\overline{H}}$ matrix in terms of the transmitter's and receiver's fields, $\overline{E}^T_i$ and $\overline{E}^R_i$, for the $i$-th mode. Using expiremental near field scans, the fields are calculated over the domain locations, $\overline{r}_j$, with $j$ swept over the entire imaging domain. The index $i$ counts measurements, which are distinguished by the excitation frequency, the control mask on the aperture, and the position of the synthetic aperture. In this sense, each new illuminating pattern simply concatenates a new row on the bottom of the matrix. This capability, and the ability to define any discretized scene space to reconstruct over, makes the formulation very general. The approach can be applied to any scene size, radiation pattern, and synthetic aperture path, making it a suitable framework to provide comparisons between the different imaging experiments.

To compute the estimated reflectivity of the scene, $\overline{f}_{est}$, from the measurements, $\overline{g}$, the matrix problem needs to be inverted. The sensing matrix need not be square and even in the case that it is, the calculation of a mathematical inverse is typically ill-advised. Instead, a common approach to solve this problem is through the matched filter solution \cite{papoulis2002probability,soumekh1999SAR}. In effect, this operation cross-correlates the measurements with the interrogating field patterns and takes the form

\begin{equation}
\label{eq:eq4}
\overline{f}_{est}=\overline{\overline{H}}^\dagger \overline{g}
\end{equation}

\noindent where $\dagger$ denotes the conjugate transpose operator. More sophisticated approaches exist, but in the case of overdetermined systems (when $\overline{\overline{H}}$ has more rows than column) this solution is fairly reliable. Regularization or iterative approaches can be employed, but these modifications require significant computational power or some prior information about the system/scene \cite{Pulido-Mancera2016c}. The added computational demands are particularly substantial in SAR environments, where the large number of synthetic aperture positions, frequency points, and tuning states, leads to an extremely large measurement set \cite{SARmanyVariables}.

A large factor in the simplicity of this model is derived from the use of the Born approximation, which linearizes the inverse problem \cite{lin1990invScatBorn}. While this model may break down in scenes with multiply-scattering features, the approach typically returns satisfactory qualitative images. The method can only return a profile of the scattering potential of the scene and is not rigorous enough to identify quantitative material characteristics. This level of reconstruction is acceptable for SAR in the discussed applications and captures the physics of the imaging process, making it the method of choice for our implementation.

\section{Lengthwise-Motional Imaging}
\label{sec:sec4}

Given the electrically-large size of the aperture in one direction, it can be deployed in SAR imaging systems in two different movement styles: \emph{lengthwise motion}, in which the aperture is translated along its long direction, and \emph{crosswise motion}, in which the aperture is translated transverse to its length. The imaging style using lengthwise motion is depicted in Fig. \ref{fig:f1} and Fig. \ref{fig:f4}. Operating under this scenario, we will begin by demonstrating stripmap and spotlight SAR imaging using the dynamic aperture. Subsequently, we illuminate the scene with random patterns, operating in a modality which we refer to as \emph{diverse pattern stripmap}. For completeness, a variety of metallic objects have been imaged to showcase the independence of the method on scene composition.

\subsection{Demonstrated Imaging Modalities}
\label{ssec:ssecA}

Stripmap and spotlight SAR are easily implemented with the steered beams demonstrated in section \ref{sec:sec2}. The pros and cons of each case are well-documented throughout the literature \cite{soumekh1999SAR,curlander1991SAR} and each mode is readily exercised with traditional hardware platforms such as phased array and reflector dish systems. Here, our goal is not to extend these imaging scenarios beyond what is already reported, but to demonstrate similar imaging performance using the low-profile and economical aperture layer made accessible by dynamic metasurfaces.

\begin{figure}
	\centering
	\begin{tikzpicture}
	\node[anchor=south west,inner sep=0] (image) at (0,0) {\includegraphics[width=3.4in]{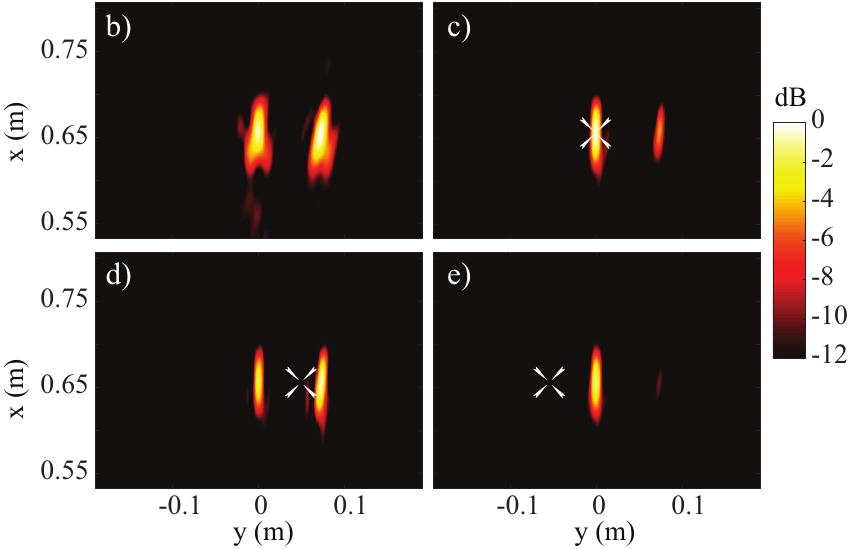}};
	
	\begin{scope}[x={(image.south east)},y={(image.north west)}]
	\node[anchor=south west,inner sep=0] (image) at (0.27,1.01) {\includegraphics[width=1.6in]{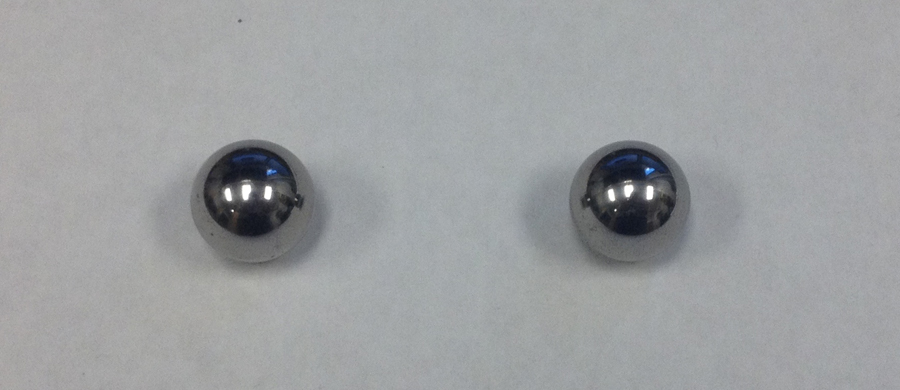}};

	\node at (0.295,1.275) {a)};

	\end{scope}
	\end{tikzpicture}
	\caption{\label{fig:f5}The objects, two metallic spheres, and their reconstructions using b) stripmap, c) centrally-aimed spotlight, d) right-aimed spotlight, e) left-aimed spotlight. The location of the center of the spotlight is denoted by a reticle.}
\end{figure}

For the stripmap case performed here, the beam is always kept at the antenna's broadside, while in the spotlight case the beam is aimed at a specific point. The beamwidth in various SAR applications arises as a trade-off with the signal-to-noise ratio (SNR) achievable in the imaging environment. For the proof-of-concept experiments reported here, measurements are completed in an anechoic chamber which ensures decent SNR---in this environment, a wider beamwidth could be used in the stripmap case to increase resolution or in the spotlight case to enlarge the scene size. However, in most SAR applications the target domain is at a significant distance and SNR can be a serious concern \cite{brown1969SARintroduction}. Thus, the narrow beamwidth used is more representative of a realistic system. Alternatively, the antenna could be made to have an adjustable beamwidth so that the SNR and resolution could be balanced on an application-to-application basis but this prospect, while possible with dynamic metasurfaces, is beyond the scope of the present work. 

For both stripmap and spotlight, the bandwidth of available beams stretches from \SI{17.50}{\giga\hertz} to \SI{20.38}{\giga\hertz} and this range is used for each imaging case. The synthetic aperture is \SI{60}{\centi\meter} long and is sampled at intervals of ${\sim}\lambda/3$ ($\Delta y_s =$ \SI{5}{\milli\meter}); the bandwidth is sampled at intervals of \SI{90}{\mega\hertz}. To illustrate some of the traits of the stripmap and spotlight modalities we reconstructed a simple scene consisting of two spheres (diameter = \SI{1.5}{\centi\meter}) separated by \SI{8}{\centi\meter}. The results for stripmap and for centrally-aimed spotlight are shown in Fig. \ref{fig:f5}b and c, respectively.

First, it is seen that the spotlight case has better SNR and resolution than the stripmap case. This is expected as the stripmap resolution is limited by the beamwidth. However, it is also worth noting that the object to the right of the scene appears slightly faded compared to the centered object in the spotlight case. This is the result of the small region of interest used in spotlight and is caused by the direct focus on a single point of interest in the scene. In fact, the spotlight can be pointed at other locations to shift emphasis towards different positions. This case is demonstrated in Fig. \ref{fig:f5}d, where the beam is directed toward the right object.  While a higher quality image is obtained in Fig. \ref{fig:f5}c, these images also expose a weakness in the spotlight modality which arises in the event that objects lie outside the area of interest. This problematic case is shown in Fig. \ref{fig:f5}d where the beam is pointed to the left of the objects and the sphere placed to the right of the scene almost completely disappears from the reconstruction. It is worth emphasizing that the tendencies observed here are inherent to stripmap and spotlight imaging and are not the result of using the dynamic metasurface at hand. From these images, it can be concluded that the dynamic metasurface aperture platform can perform stripmap and spotlight SAR with a simple architecture and without reliance on phase shifters or mechanical rotation.

\begin{figure}
	\centering
	\begin{tikzpicture}
	\node[anchor=south west,inner sep=0] (image) at (0,0) {\includegraphics[width=3.4in]{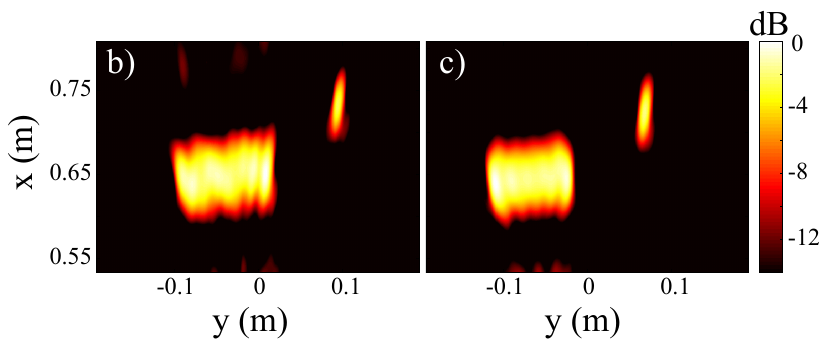}};
	
	\begin{scope}[x={(image.south east)},y={(image.north west)}]
	\node[anchor=south west,inner sep=0] (image) at (0.29,0.91) {\includegraphics[width=1.6in]{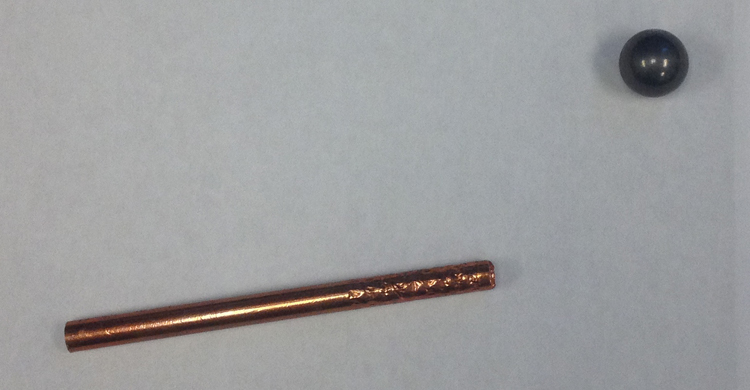}};
	
	\node at (0.315,1.43) {a)};
	
	\end{scope}
	\end{tikzpicture}
	\caption{\label{fig:f6}a) The objects, a metallic pen and a sphere, and their diverse pattern stripmap images from b) experiment and c) simulation}
\end{figure}

As mentioned earlier, we expect diverse pattern stripmap to mitigate the intrinsic limitations of stripmap and spotlight imaging. To experimentally validate the feasibility of this imaging approach, described visually in Fig. \ref{fig:f1}c, we demonstrate the ability to reconstruct a complicated scene, shown in Fig. \ref{fig:f6}a. We use 10 different masks for each frequency and aperture location (with the same frequency vector and aperture sampling as the previous measurements). In total 61,710 measurements are taken and used to reconstruct the scene. The results for this demonstration are shown in Fig. \ref{fig:f6}. The metallic targets are a \SI{2}{\centi\meter} diameter ball and a \SI{10}{\centi\meter} long, \SI{1}{\centi\meter} diameter cylinder. These objects vary in response in that the cylinder is a large, continuous object with specular reflections while the ball is relatively small and scatters in all directions. Both objects appear starkly (with the cylinder being slightly brighter due to its nearer position) and stand out well above the background noise.

To ensure that the above performance is in line with expectations, a simulation of the ideal scenario has been completed using the formulation outlined in \cite{lipworth2013JOSA}. The aperture's experimental characterization is used to propagate the fields to the scene and the reflective objects are represented as collections of point scatterers. The simulated backscattered signal is computed with Eq. \ref{eq:eq3} by calculating the transmitted fields, the scattering event, and the backprojected receive fields. The objects are modeled to be the same as those used in the experiment. As can be seen in the reconstruction, shown in Fig. \ref{fig:f6}c, excellent agreement is found between the experiment and the best case simulation. The appearance of the cylinder is similar in the two cases and the resolution, as understood from the metallic ball, is very comparable. This result hints at the importance of resolution and field of view (FOV) in SAR imaging platforms. We will quantify these properties more rigorously in section \ref{ssec:ssecC}, but first we address the sampling of the aperture.

\subsection{Spatial Sampling}
\label{ssec:ssecB}

\begin{figure}
	\centering
	\begin{tikzpicture}
	\node[anchor=south west,inner sep=0] (image) at (0,0) {\includegraphics[width=3.4in]{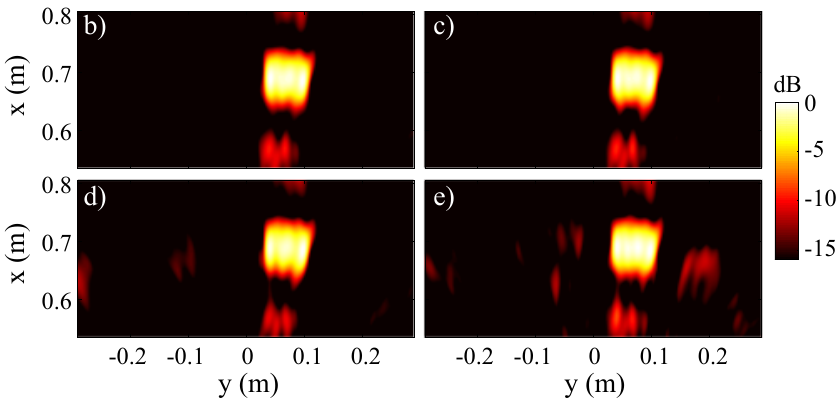}};
	
	\begin{scope}[x={(image.south east)},y={(image.north west)}]
	\node[anchor=south west,inner sep=0] (image) at (0.27,1) {\includegraphics[width=1.6in]{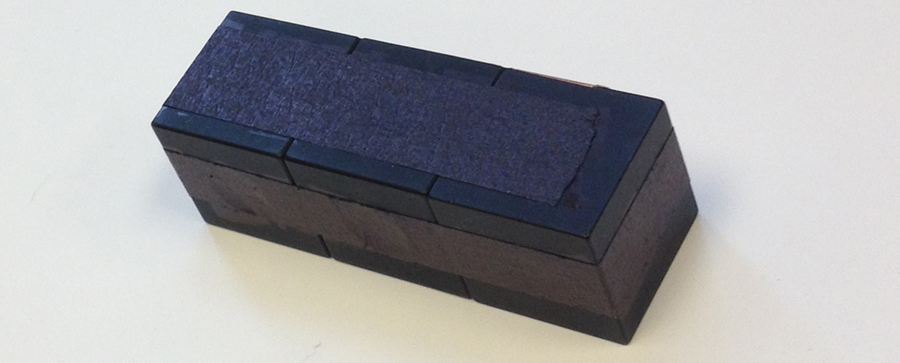}};
	
	\node at (0.295,1.335) {a)};
	
	\end{scope}
	\end{tikzpicture}
	\caption{\label{fig:f7}a) The object, a rectangular prism, and its reconstruction with various samplings of the synthetic aperture: b) \SI{0.5}{\centi\meter}, c) \SI{2}{\centi\meter}, d) \SI{3}{\centi\meter}, e) \SI{8}{\centi\meter}.}
\end{figure}

Various works have recently investigated the effects of spatial sampling in array imaging and the potential of using sparse elements with aperiodic arrangements \cite{gonzalez2014sparse,yurduseven2016Aperiodic}. Our physical aperture is electrically-large and therefore is not necessarily subject to the usual $\lambda/2$ sampling rate, but the aperture is filled inhomogeneously with fields falling off as the distance from the center increases. As such, any absolute statement about the effective aperture size and its relation to the physical aperture size is tenuous. Here, we conduct an empirical study to see how the sampling of the synthetic aperture affects the reconstruction. By sweeping the increment size ($\Delta y_s$ in Fig. \ref{fig:f4}) it can be seen how the sampling of the synthetic aperture affects reconstructions; various increments are shown in Fig. \ref{fig:f7} where an elongated rectangular prism is imaged. Within this study, the number of masks was increased for the sparser cases so that the overall number of measurements remained the same. As expected, sampling can be coarser than the typical $\lambda/2$ rate, but substantial sparsity eventually gives rise to spatial aliasing. From these results, it appears that \SI{2}{\centi\meter} (${\sim}1.25\lambda$) is a reasonable step size and as such we will retain this sampling throughout this work. 

\subsection{Comparisons with Diverse Pattern Stripmap}
\label{ssec:ssecC}

\begin{figure}
	\centering
	\begin{tikzpicture}
	\node[anchor=south west,inner sep=0] (image) at (0,0) {\includegraphics[width=3.4in]{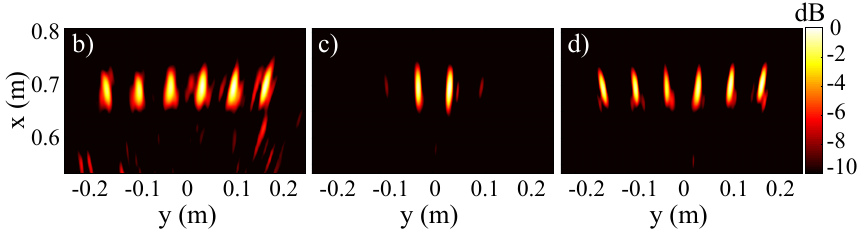}};
	
	\begin{scope}[x={(image.south east)},y={(image.north west)}]
	\node[anchor=south west,inner sep=0] (image) at (0.18,0.92) {\includegraphics[width=2.2in]{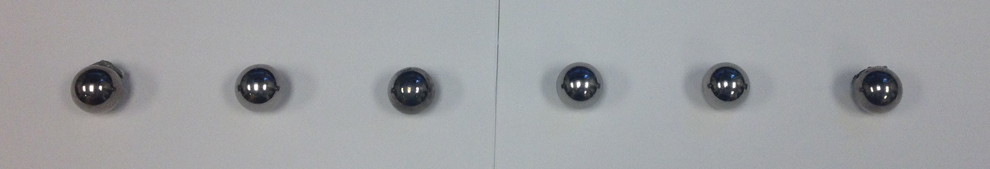}};
	
	\node at (0.2,1.24) {a)};
	
	\end{scope}
	\end{tikzpicture}
	\caption{\label{fig:f8}a) The objects, six spheres separated at \SI{3}{\centi\meter}, and the results with b) stripmap, c) spotlight, d) diverse pattern stripmap.}
\end{figure}

Having demonstrated the experimental feasibility of diverse pattern stripmap SAR, we now directly compare it to stripmap and spotlight. To quantitatively compare resolution and FOV between the conventional modes and the diverse pattern mode, we will use a large region of interest with spheres (diameter = \SI{1.5}{\centi\meter}) placed along a line of constant depth. Six spheres are included which are separated by \SI{5}{\centi\meter} and collectively span a distance of \SI{25}{\centi\meter}. The results for the three modalities are shown in Fig. \ref{fig:f8}. It is seen that diverse pattern stripmap achieves the same cross range resolution as spotlight and has the same FOV as stripmap. Meanwhile, spotlight mode struggles with the large scene size and stripmap suffers from poor resolution.

\begin{figure} 
	\centering
	\includegraphics[width=3.3in]{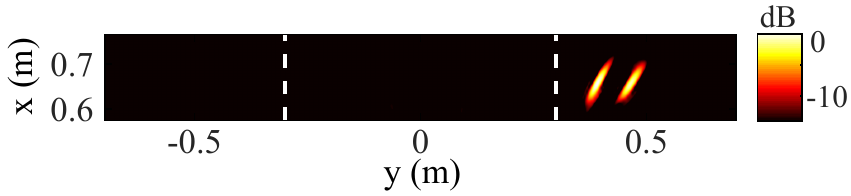}
	\caption{\label{fig:f9}Image of two spheres located outside of the synthetic aperture's transverse path (marked by dashed lines), obtained with diverse pattern stripmap.}
\end{figure}

Additionally, diverse patterns can image over a larger FOV since the radiation patterns are not confined to the boresight of the antenna. This feature is further highlighted in Fig. \ref{fig:f9} by placing a pair of spheres near the edges of the imaging domain. Note that the objects appear slanted because the object blur along the radial direction reveals the range resolution. Ultimately, a FOV of \SI{45}{\degree} beyond the edges of the synthesized aperture can be examined.

\begin{figure} 
	\centering
	\includegraphics[width=3.3in]{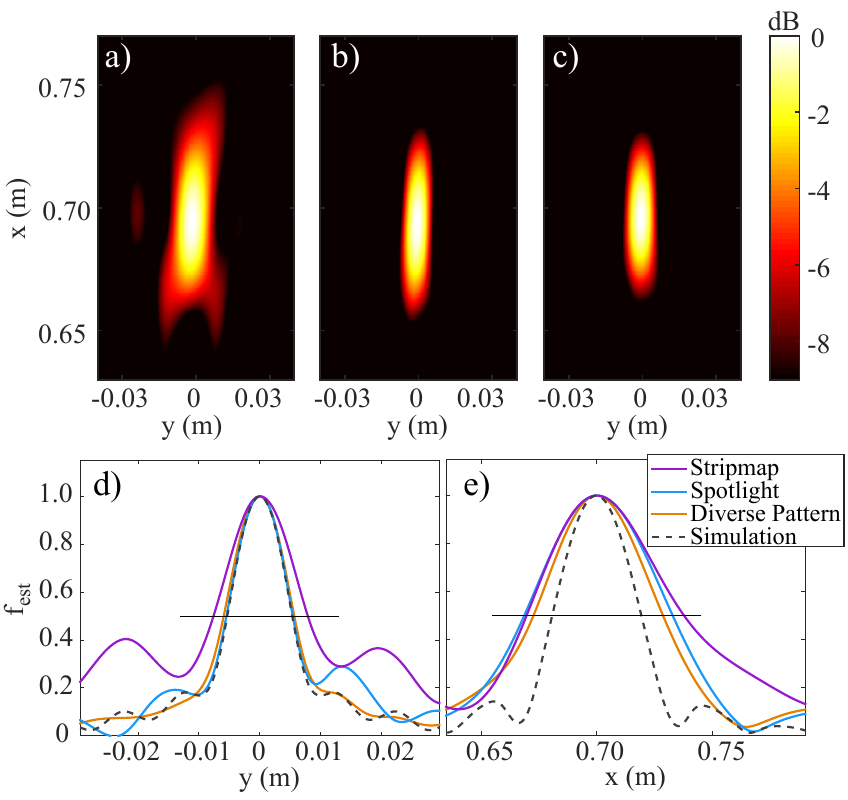}
	\caption{\label{fig:f10}Two-dimension PSFs for a) stripmap, b) spotlight, c) diverse pattern stripmap. Cross sections for the d) cross-range and e) range are also shown for the various modes and for the ideal simulation.}
\end{figure}

By conducting a point spread function (PSF) analysis the resolution can be further quantified \cite{yurduseven2015PSF}. In this experiment we reconstruct a subwavelength spherical scatterer and monitor the full width at half maximum (FWHM) of the reconstructed images. For all three imaging methods, aerial-view reconstructions and cross sections along the main axes are plotted in Fig. \ref{fig:f10}. In the range direction, the reconstructions have comparable behavior because the same bandwidth is utilized. For the cross-range resolution, $\delta_y$, spotlight and diverse pattern modes have virtually equivalent resolution (\SI{1.1}{\centi\meter}) whereas the stripmap performs slightly worse (\SI{1.6}{\centi\meter}) owing to its beamwidth limitation. Since we are operating in the radiative near-field and the synthetic aperture size is constrained by the physical hardware, a straightforward formula for diffraction-limited resolution is not readily available. Instead, we conduct a numerical simulation to determine how these results compare to the ideal case. To this end, a dipole antenna is translated through the same synthetic aperture with the same sampling and bandwidth characteristics. The result is plotted as a dashed line in Fig. \ref{fig:f10}. Evidently, the ideal cross-range resolution is in excellent agreement with the experimental ones obtained using spotlight and diverse pattern stripmap. There is noticeable discrepancy in the range result however; this is due to the unleveled power spectrum emitted from the metasurface antenna, which has higher radiated power at lower frequencies \cite{sleasmanMarathon}.

The results presented in Figs. \ref{fig:f8}-\ref{fig:f10} showcase the advantages offered by using diverse pattern stripmap. In particular, diverse pattern stripmap obtains similar resolution to the spotlight modality while accessing a larger FOV than the stripmap modality. While all these experiments have been conducted using the dynamic metasurface, it should be emphasized that the limitations of spotlight and stripmap imaging are well-documented are not result of using the dynamic metasurface.

It should be noted that the diverse pattern modality incurs a hit in SNR due to the fact that the energy is spread across an overall wider area. This factor is not encountered as a practical difficulty in the present laboratory environment. Additionally, since tuning through 10 masks in our experimental configuration effectively means a longer integration time at each position, we also average the measurements in the spotlight/stripmap cases 10 times to ensure fair comparisons. Further application-specific limitations will be investigated in a future work from a more technical perspective; our goal in this paper is to explore preliminary conceptual trade-offs in terms of image quality.

\section{Crosswise-Motional Imaging}
\label{sec:sec5}

\begin{figure} 
	\centering
	\includegraphics[width=2.3in]{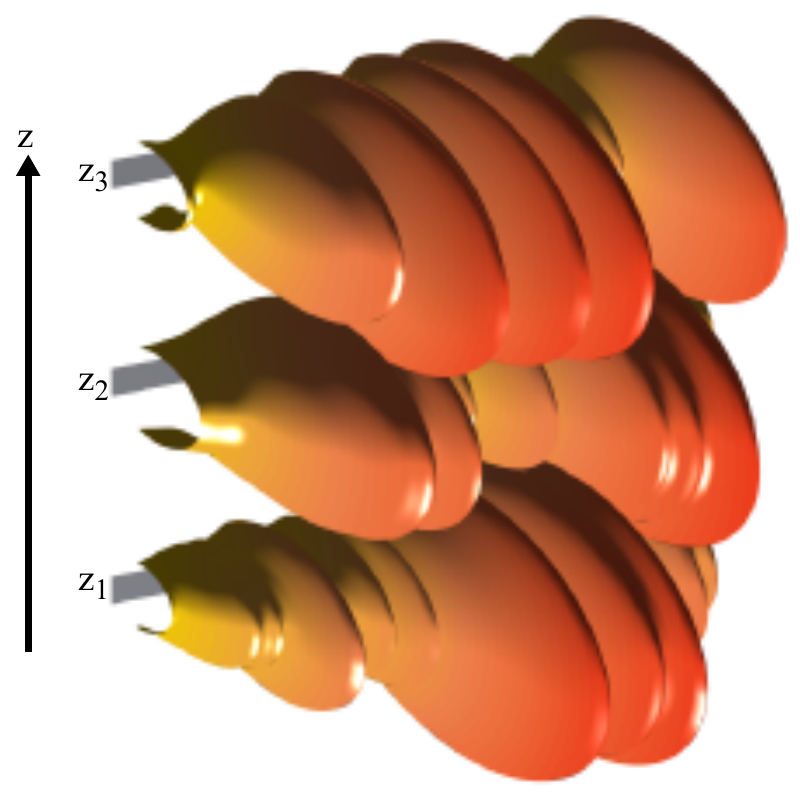}
	\caption{\label{fig:f11}The aperture shifting in the crosswise direction, operating in diverse pattern modality.}
\end{figure}

Consistent with existing SAR imaging, the previous section outlined how, using a combination of mechanical motion and frequency-swept measurements, 2D information can be obtained.  This type of imaging uses the mechanical motion to obtain cross-range information and uses the frequency measurements to obtain range information.  In this section, we describe how we can leverage the ability of the dynamic metasurface to radiate different patterns to probe the cross range spatial content of a scene without mechanical motion.  By using mechanical motion, structured radiation patterns, and frequency-swept measurements, information can be obtained in all three dimensions to perform volumetric imaging. A schematic of this mode of operation is shown in Fig. \ref{fig:f11}. Note that the aperture is translated along z (granting $\delta_z$), while its patterns vary in y (granting $\delta_y$), and has range information in x from frequency (granting $\delta_x$). Obviously, resolutions are not entirely decoupled, but this description gives an intuitive perspective on how the different features contribute to each resolution. 

\begin{figure} 
	\centering
	\includegraphics[width=3.15in]{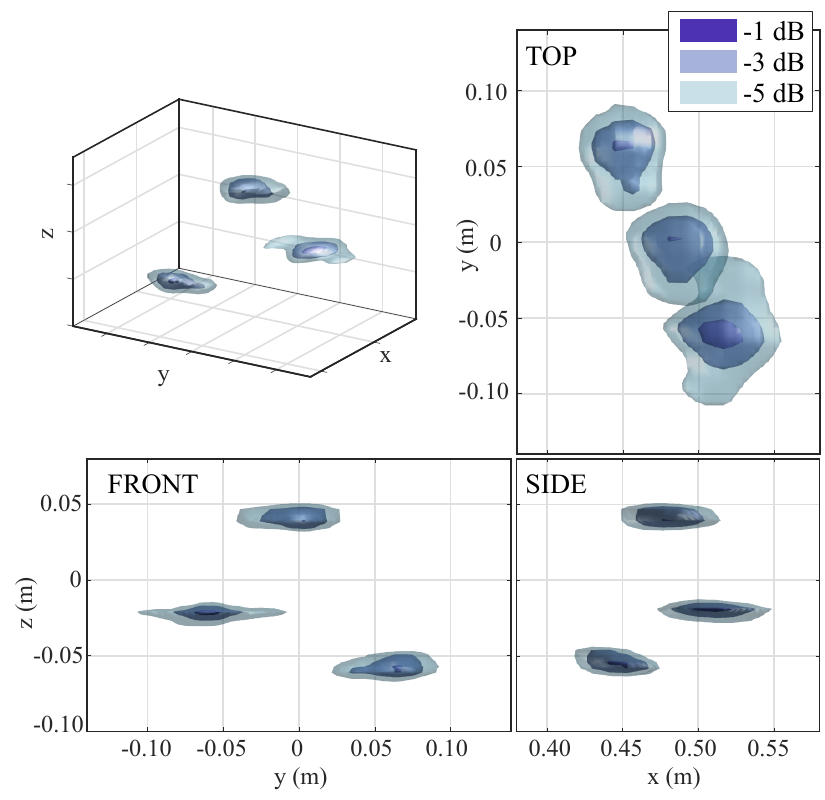}
	\caption{\label{fig:f12}Diverse pattern 3D imaging result for three spheres.}
\end{figure}

\begin{figure} 
	\centering
	\includegraphics[width=3.15in]{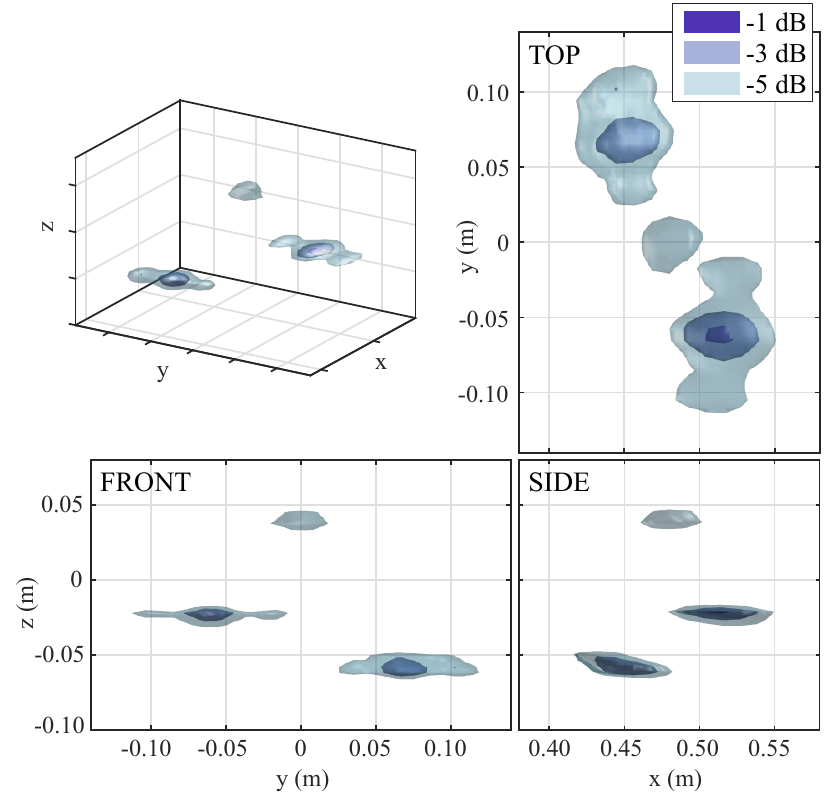}
	\caption{\label{fig:f13}Scan SAR 3D imaging result for three spheres.}
\end{figure}

The diverse pattern modality described above is naturally adapted to this motional framework.   Alternatively, the spotlight ability can be used to scan across the y direction---this is the scan SAR imaging method \cite{scanSAR1981,scanSAR1996}. In scan SAR, the angular sweep ideally occurs for each z position and covers all frequency points. In this work, we step the angle by \SI{4}{\degree} between \SI{\pm 20}{\degree}. In this sense, every angle must be probed independently and thus the beam must traverse the entire scene. This is in contrast to the diverse pattern mode which multiplexes the y spatial information and relies on post-processing to retrieve the image. In a sense, diverse pattern stripmap sends multiple beams simultaneously and then demultiplexes the spatial components in calculation.

\begin{figure}
	\begin{tikzpicture}
	\node[anchor=south west,inner sep=0] (image) at (0,0) {\includegraphics[width=1.9in]{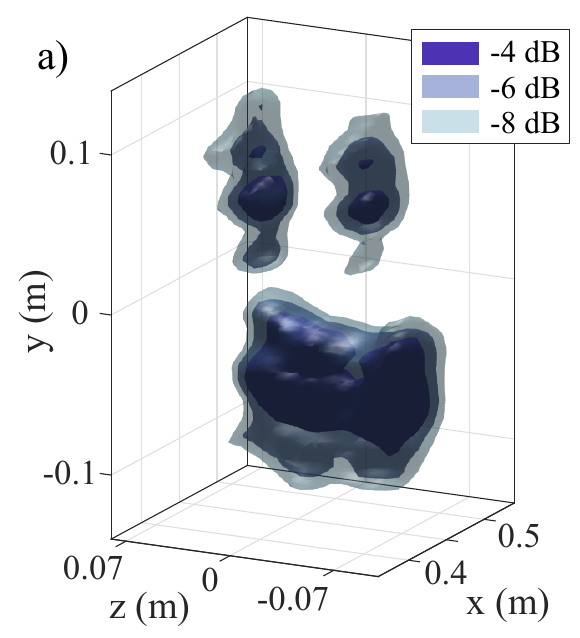}};
	
	\begin{scope}[x={(image.south east)},y={(image.north west)}]
	\node[anchor=south west,inner sep=0] (image) at (1.02,0.1) {\includegraphics[width=1.3in]{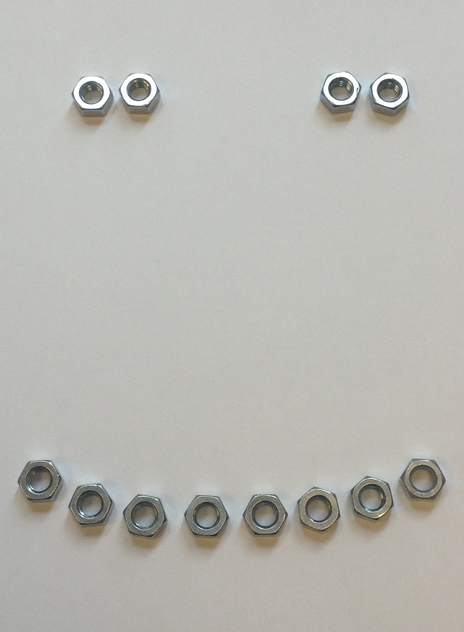}};
	
	\node at (1.07,0.88) {b)};
	
	\end{scope}
	\end{tikzpicture}
	\caption{\label{fig:f14}Diverse pattern 3D imaging result for b) a complex target that is a smiley face composed of nuts.}
\end{figure}

With the angle span and sampling for the scan SAR case, it is seen that 11 angular measurements must be taken at each z position, for the entire frequency sweep. While it seems fair that the diverse pattern case uses a similar number of measurements (i.e. 11 masks), we know that the aperture can image with fewer measurements in this modality \cite{sleasman2015DyAp1,sleasmanMarathon}. As such, we use a new set of 3 masks at each z position. The synthetic aperture is \SI{32.25}{\centi\meter} long and is sampled at $\Delta z_s =$ \SI{7.5}{\milli\meter} and the same frequency vector is retained.

We begin by demonstrating each of these methods with a set of simple objects. Three spheres are distributed in the three dimensions and imaged with both diverse pattern and scan SAR modalities. We have plotted the projections in the three different dimensions in Fig. \ref{fig:f12}, for the diverse pattern case, and Fig. \ref{fig:f13}, for the scan SAR case. It is seen that the spheres are easily resolved in both cases. Each approach here has the same bandwidth and frequency sampling, but the diverse pattern approach uses 3 measurements per position as compared to the 11 angular measurements for scan SAR. The angular sampling or span can be decreased to reduce the acquisition time, but either of these adjustments could result in severe deterioration of reconstructed images.  The angular sampling used here is not optimal, only an initial demonstration of the aperture's potential.

\begin{figure}
	\centering
	\begin{tikzpicture}
	\node[anchor=south west,inner sep=0] (image) at (0,0) {\includegraphics[width=3.4in]{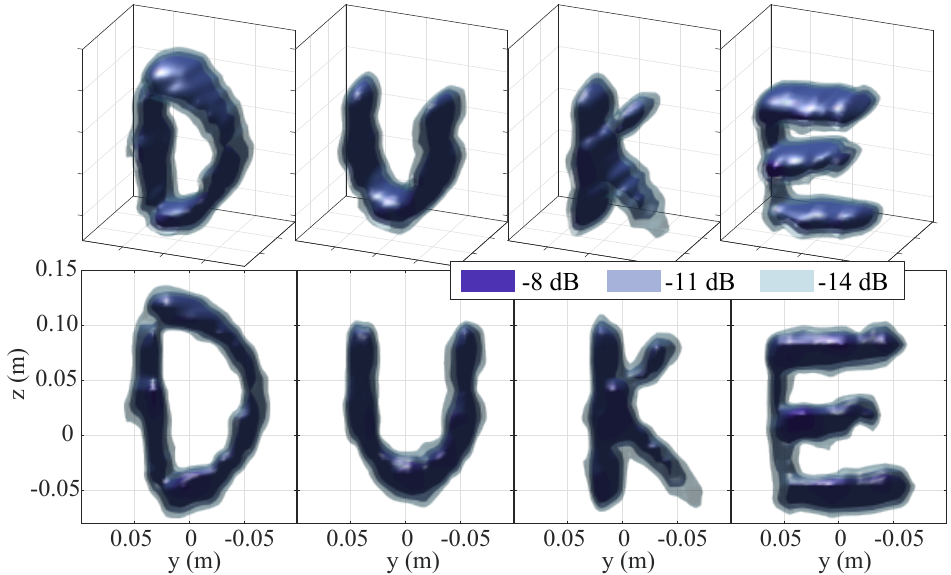}};
	
	\begin{scope}[x={(image.south east)},y={(image.north west)}]
	\node[anchor=south west,inner sep=0] (image) at (0.09,1.01) {\includegraphics[width=3.in]{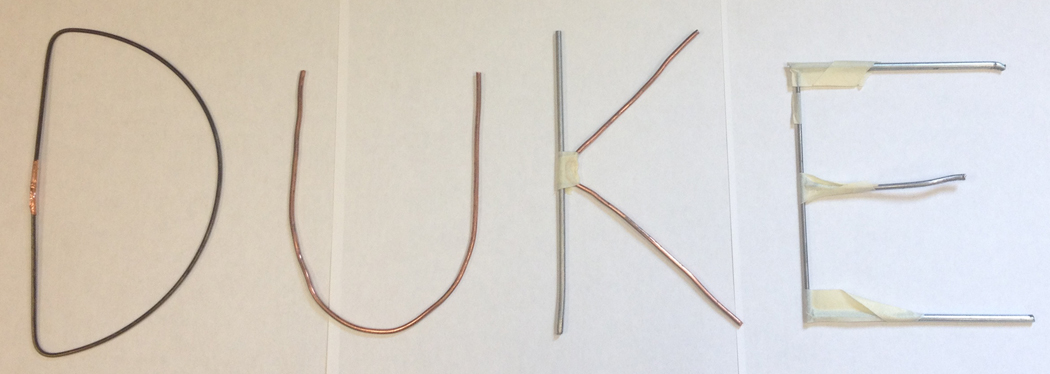}};
	
	\end{scope}
	\end{tikzpicture}
	\caption{\label{fig:f15}Meandering imaging result for the letters spelling DUKE; isometric and front-facing results are shown.}
\end{figure}

We have also reconstructed a more complex object with the diverse pattern stripmap platform. Figure \ref{fig:f14} shows a smiley face which is composed of \SI{8}{\milli\meter} diameter nuts and has more distinct features. Owing to the length of the synthetic aperture and the length of the physical aperture, it is seen in Fig. \ref{fig:f12}-\ref{fig:f14} that the resolution is higher in the z direction than in the y direction. However, in all cases the quality of reconstruction is evidence of the aperture's utility in 3D imaging.

As a last experiment we translate the aperture in both $y$ and $z$ directions. In this case, the aperture moves in a meandered trajectory to span \SI{24}{\centi\meter} in the z direction and \SI{20}{\centi\meter} in the y direction. Step sizes of \SI{20}{\milli\meter} are taken in y while step sizes of \SI{10}{\milli\meter} are taken in z. The same frequencies are used and 3 masks are employed at each location. With the additional size in the y direction, the improved resolution enables images of more complex objects. Here, we have imaged cylindrical wires (diameter = \SI{3}{\milli\meter}) bent into the shapes of letters. Figure \ref{fig:f15} shows separate reconstructions of D, U, K, and E.

\section{Conclusion}
\label{sec:sec6}

We have demonstrated high-quality synthetic aperture radar imaging from a dynamic metasurface antenna platform. The aperture possesses several advantages in terms of form factor, power consumption, and cost when compared to traditional systems. Spotlight and stripmap SAR have been shown using steered beams enabled by the metasurface antenna. Additionally, a third modality, diverse pattern stripmap, has been introduced and proven to have favorable performance. Resolution and field of view studies have been conducted for the three methods and the various trade-offs between the modes have been investigated. In particular, it was shown that diverse pattern stripmap was able to achieve the same resolution as the spotlight case with a larger field of view than either conventional mode.

In addition to the 2D imaging case that is frequently seen in airborne and spaceborne SAR, we have also investigated 3D imaging by translating the aperture along a path perpendicular to its length. This approach, distinct from interferometric SAR, provides full volumetric reconstructions by taking advantage of the frequency bandwidth, physical aperture size, and synthetic path. Scanned beams and diverse patterns have been shown in this movement style, with the diverse pattern case providing a multiplexing advantage to cut down on the number of measurements. Finally, the aperture has been translated through a meandering path to provide volumetric images with high resolution in all three dimensions.

The aperture may find use in a variety of SAR-related applications. Historically, SAR has been employed from aircraft and satellite systems but the technique has increasingly been applied in cases where the standoff distance is significantly shorter. Holographic imaging systems \cite{nikolovaMicrowaveHolo,brady2009CSholo} for security screening \cite{sleasmanMarathon,hunt2014metaimager,sheen2001threatimaging,alvarez2015_3dCS,yarovoySparseSAR} and feature-specific imaging \cite{neifeld2003featspecimag,gehmFeatSpecific} are of increasing interest and the dynamic metasurface aperture is posed to make important contributions across the entire field of microwave sensing.

\section*{Acknowledgment}
This work was supported by the Air Force Office of Scientific Research (AFOSR, Grant No. FA9550-12-1-0491).

\bibliographystyle{IEEEtran}
\bibliography{sleasman_expSAR}

\begin{thebibliography}{10}
\providecommand{\url}[1]{#1}
\csname url@samestyle\endcsname
\providecommand{\newblock}{\relax}
\providecommand{\bibinfo}[2]{#2}
\providecommand{\BIBentrySTDinterwordspacing}{\spaceskip=0pt\relax}
\providecommand{\BIBentryALTinterwordstretchfactor}{4}
\providecommand{\BIBentryALTinterwordspacing}{\spaceskip=\fontdimen2\font plus
\BIBentryALTinterwordstretchfactor\fontdimen3\font minus
  \fontdimen4\font\relax}
\providecommand{\BIBforeignlanguage}[2]{{%
\expandafter\ifx\csname l@#1\endcsname\relax
\typeout{** WARNING: IEEEtran.bst: No hyphenation pattern has been}%
\typeout{** loaded for the language `#1'. Using the pattern for}%
\typeout{** the default language instead.}%
\else
\language=\csname l@#1\endcsname
\fi
#2}}
\providecommand{\BIBdecl}{\relax}
\BIBdecl

\bibitem{curlander1991SAR}
J.~C. Curlander and R.~N. McDonough, \emph{Synthetic aperture radar}.\hskip 1em
  plus 0.5em minus 0.4em\relax John Wiley \& Sons, 1991.

\bibitem{soumekh1999SAR}
M.~Soumekh, \emph{Synthetic aperture radar signal processing}.\hskip 1em plus
  0.5em minus 0.4em\relax New York: Wiley, 1999.

\bibitem{brown1969SARintroduction}
W.~M. Brown and L.~J. Porcello, ``An introduction to synthetic-aperture
  radar,'' \emph{IEEE spectrum}, vol.~6, no.~9, pp. 52--62, 1969.

\bibitem{Pulido-Mancera2016c}
L.~Pulido-Mancera, T.~Fromenteze, T.~Sleasman, M.~Boyarsky, M.~F. Imani,
  M.~Reynolds, and D.~Smith, ``Application of range migration algorithms to
  imaging with a dynamic metasurface antenna,'' \emph{JOSA B}, vol.~33, no.~10,
  pp. 2082--2092, 2016.

\bibitem{soumekh1994fourierarray}
M.~Soumekh, \emph{Fourier array imaging}.\hskip 1em plus 0.5em minus
  0.4em\relax Prentice-Hall, Inc., 1994.

\bibitem{yarovoySparseSAR}
X.~Zhuge and A.~G. Yarovoy, ``A sparse aperture mimo-sar-based uwb imaging
  system for concealed weapon detection,'' \emph{IEEE Transactions on
  Geoscience and Remote Sensing}, vol.~49, no.~1, pp. 509--518, Jan 2011.

\bibitem{zhuge2012three}
X.~Zhuge and A.~Yarovoy, ``Three-dimensional near-field mimo array imaging
  using range migration techniques,'' \emph{IEEE Transactions on Image
  Processing}, vol.~21, no.~6, pp. 3026--3033, 2012.

\bibitem{sheen2001threatimaging}
D.~M. Sheen, D.~L. McMakin, and T.~E. Hall, ``Three-dimensional millimeter-wave
  imaging for concealed weapon detection,'' \emph{IEEE Trans. Microw. Theory
  Techn.}, vol.~49, no.~9, pp. 1581--1592, 2001.

\bibitem{alvarez2015_3dCS}
Y.~Alvarez, Y.~Rodriguez-Vaqueiro, B.~Gonzalez-Valdes, C.~Rappaport,
  F.~Las-Heras, and J.~Martinez-Lorenzo, ``Three-dimensional compressed
  sensing-based millimeter-wave imaging,'' \emph{IEEE Trans. Antennas Propag.},
  vol.~63, no.~12, pp. 5868--5873, Dec 2015.

\bibitem{gonzalez2014sparse}
B.~Gonzalez-Valdes, G.~Allan, Y.~Rodriguez-Vaqueiro, Y.~Alvarez,
  S.~Mantzavinos, M.~Nickerson, B.~Berkowitz, J.~A. Martinez-Lorenzo,
  F.~Las-Heras, and C.~M. Rappaport, ``Sparse array optimization using
  simulated annealing and compressed sensing for near-field millimeter wave
  imaging,'' \emph{Antennas and Propagation, IEEE Transactions on}, vol.~62,
  no.~4, pp. 1716--1722, 2014.

\bibitem{ahmed2011Imaging}
S.~S. Ahmed, A.~Schiessl, and L.~P. Schmidt, ``A novel fully electronic active
  real-time imager based on a planar multistatic sparse array,'' \emph{IEEE
  Transactions on Microwave Theory and Techniques}, vol.~59, no.~12, pp.
  3567--3576, Dec 2011.

\bibitem{ahmed2012Imaging}
S.~S. Ahmed, A.~Schiessl, F.~Gumbmann, M.~Tiebout, S.~Methfessel, and L.~P.
  Schmidt, ``Advanced microwave imaging,'' \emph{IEEE Microwave Magazine},
  vol.~13, no.~6, pp. 26--43, Sept 2012.

\bibitem{sarabandi1992EarthImage}
Y.~Oh, K.~Sarabandi, and F.~T. Ulaby, ``An empirical model and an inversion
  technique for radar scattering from bare soil surfaces,'' \emph{IEEE
  Transactions on Geoscience and Remote Sensing}, vol.~30, no.~2, pp. 370--381,
  Mar 1992.

\bibitem{treuhaft1996vegetation}
R.~N. Treuhaft, S.~N. Madsen, M.~Moghaddam, and J.~J. Zyl, ``Vegetation
  characteristics and underlying topography from interferometric radar,''
  \emph{Radio Science}, vol.~31, no.~6, pp. 1449--1485, 1996.

\bibitem{dehmollaian2008throughwall}
M.~Dehmollaian and K.~Sarabandi, ``Refocusing through building walls using
  synthetic aperture radar,'' \emph{Geoscience and Remote Sensing, IEEE
  Transactions on}, vol.~46, no.~6, pp. 1589--1599, 2008.

\bibitem{huang2010throughwall}
Q.~Huang, L.~Qu, B.~Wu, and G.~Fang, ``Uwb through-wall imaging based on
  compressive sensing,'' \emph{IEEE Transactions on Geoscience and Remote
  Sensing}, vol.~48, no.~3, pp. 1408--1415, March 2010.

\bibitem{jakowatz2012spotlight}
C.~V. Jakowatz, D.~E. Wahl, P.~H. Eichel, D.~C. Ghiglia, and P.~A. Thompson,
  \emph{Spotlight-Mode Synthetic Aperture Radar: A Signal Processing Approach:
  A Signal Processing Approach}.\hskip 1em plus 0.5em minus 0.4em\relax
  Springer Science \& Business Media, 2012.

\bibitem{hansen2009phasedarray}
R.~C. Hansen, \emph{Phased array antennas}.\hskip 1em plus 0.5em minus
  0.4em\relax John Wiley \& Sons, 2009, vol. 213.

\bibitem{johnson2015sidelobeCancel}
M.~C. Johnson, S.~L. Brunton, N.~B. Kundtz, and J.~N. Kutz, ``Sidelobe
  canceling for reconfigurable holographic metamaterial antenna,'' \emph{IEEE
  Transactions on Antennas and Propagation}, vol.~63, no.~4, pp. 1881--1886,
  April 2015.

\bibitem{patel2014ESA}
D.~J. Gregoire, J.~S. Colburn, A.~M. Patel, R.~Quarfoth, and D.~Sievenpiper,
  ``A low profile electronically-steerable artificial-impedance-surface
  antenna,'' in \emph{Electromagnetics in Advanced Applications (ICEAA), 2014
  International Conference on}, Aug 2014, pp. 477--479.

\bibitem{patel2015ESA}
D.~J. Gregoire, A.~Patel, and R.~Quarfoth, ``A design for an
  electronically-steerable holographic antenna with polarization control,'' in
  \emph{2015 IEEE International Symposium on Antennas and Propagation USNC/URSI
  National Radio Science Meeting}, July 2015, pp. 2203--2204.

\bibitem{lim2004CRLHsteer}
S.~Lim, C.~Caloz, and T.~Itoh, ``Metamaterial-based electronically controlled
  transmission-line structure as a novel leaky-wave antenna with tunable
  radiation angle and beamwidth,'' \emph{IEEE Transactions on Microwave Theory
  and Techniques}, vol.~52, no.~12, pp. 2678--2690, 2004.

\bibitem{hunt2013science}
J.~Hunt, T.~Driscoll, A.~Mrozack, G.~Lipworth, M.~Reynolds, D.~Brady, and D.~R.
  Smith, ``Metamaterial apertures for computational imaging,'' \emph{Science},
  vol. 339, no. 6117, pp. 310--313, 2013.

\bibitem{landy2013WGhomogenization}
N.~Landy, J.~Hunt, and D.~R. Smith, ``Homogenization analysis of complementary
  waveguide metamaterials,'' \emph{Photonics and Nanostructures-Fundamentals
  and Applications}, vol.~11, no.~4, pp. 453--467, 2013.

\bibitem{sleasman2015DyAp1}
T.~Sleasman, M.~F.~Imani, J.~N. Gollub, and D.~R. Smith, ``Dynamic metamaterial
  aperture for microwave imaging,'' \emph{Appl. Phys. Lett.}, vol. 107, no.~20,
  2015.

\bibitem{jammingSAR}
L.~Rosenberg and D.~Gray, ``Anti-jamming techniques for multichannel sar
  imaging,'' \emph{IEE Proceedings - Radar, Sonar and Navigation}, vol. 153,
  no.~3, pp. 234--242, June 2006.

\bibitem{jammingSAR2}
A.~S. Paine, ``An adaptive beamforming technique for countering synthetic
  aperture radar (sar) jamming threats,'' in \emph{2007 IEEE Radar Conference},
  April 2007, pp. 630--634.

\bibitem{mimosar}
G.~Krieger, ``Mimo-sar: Opportunities and pitfalls,'' \emph{IEEE Transactions
  on Geoscience and Remote Sensing}, vol.~52, no.~5, pp. 2628--2645, May 2014.

\bibitem{oliner1993leaky}
A.~A. Oliner and D.~R. Jackson, ``Leaky-wave antennas,'' \emph{Antenna
  Engineering Handbook}, vol.~4, p.~12, 1993.

\bibitem{hunt2014metaimager}
J.~Hunt, J.~Gollub, T.~Driscoll, G.~Lipworth, A.~Mrozack, M.~S. Reynolds, D.~J.
  Brady, and D.~R. Smith, ``Metamaterial microwave holographic imaging
  system,'' \emph{J. Opt. Soc. Amer. A}, vol.~31, no.~10, pp. 2109--2119, 2014.

\bibitem{sleasmanMarathon}
T.~Sleasman, M.~Boyarsky, M.~F.~Imani, J.~N. Gollub, and D.~R. Smith, ``Design
  considerations for a dynamic metamaterial aperture for computational imaging
  at microwave frequencies,'' \emph{J. Opt. Soc. Amer. B}, vol.~33, no.~6, pp.
  1098--1111, Jun 2016.

\bibitem{brady2009CSholo}
D.~J. Brady, K.~Choi, D.~L. Marks, R.~Horisaki, and S.~Lim, ``Compressive
  holography,'' \emph{Opt. Exp.}, vol.~17, no.~15, pp. 13\,040--13\,049, 2009.

\bibitem{brady2009optical}
D.~J. Brady, \emph{Optical imaging and spectroscopy}.\hskip 1em plus 0.5em
  minus 0.4em\relax John Wiley \& Sons, 2009.

\bibitem{donoho2006CS}
D.~L. Donoho, ``Compressed sensing,'' \emph{Information Theory, IEEE
  Transactions on}, vol.~52, no.~4, pp. 1289--1306, 2006.

\bibitem{watts2014THzCS}
C.~M. Watts, D.~Shrekenhamer, J.~Montoya, G.~Lipworth, J.~Hunt, T.~Sleasman,
  S.~Krishna, D.~R. Smith, and W.~J. Padilla, ``Terahertz compressive imaging
  with metamaterial spatial light modulators,'' \emph{Nature Photonics}, 2014.

\bibitem{Caorsi2000compImag}
S.~Caorsi, A.~Massa, and M.~Pastorino, ``A computational technique based on a
  real-coded genetic algorithm for microwave imaging purposes,'' \emph{IEEE
  Transactions on Geoscience and Remote Sensing}, vol.~38, no.~4, pp.
  1697--1708, Jul 2000.

\bibitem{fenimore1978codedaperture}
E.~E. Fenimore and T.~Cannon, ``Coded aperture imaging with uniformly redundant
  arrays,'' \emph{Appl. Opt.}, vol.~17, no.~3, pp. 337--347, 1978.

\bibitem{hand2008cELC}
T.~H. Hand, J.~Gollub, S.~Sajuyigbe, D.~R. Smith, and S.~A. Cummer,
  ``Characterization of complementary electric field coupled resonant
  surfaces,'' \emph{Applied Physics Letters}, vol.~93, no.~21, p. 212504, 2008.

\bibitem{sleasman2015element}
T.~Sleasman, M.~F.~Imani, W.~Xu, J.~Hunt, T.~Driscoll, M.~S. Reynolds, and
  D.~R. Smith, ``Waveguide-fed tunable metamaterial element for dynamic
  apertures,'' \emph{IEEE Antennas and Wireless Propag. Lett.}, vol.~15, pp.
  606--609, 2016.

\bibitem{balanis2005antenna}
C.~A. Balanis, \emph{Antenna theory: analysis and design}.\hskip 1em plus 0.5em
  minus 0.4em\relax John Wiley \& Sons, 2005, vol.~1.

\bibitem{goodmanFourierOptics}
J.~W. Goodman, \emph{Introduction to Fourier optics}.\hskip 1em plus 0.5em
  minus 0.4em\relax Roberts and Company Publishers, 2005.

\bibitem{johnson2014DDA_ApOp}
M.~Johnson, P.~Bowen, N.~Kundtz, and A.~Bily, ``Discrete-dipole approximation
  model for control and optimization of a holographic metamaterial antenna,''
  \emph{Applied optics}, vol.~53, no.~25, pp. 5791--5799, 2014.

\bibitem{Pulido-Mancera2016}
L.~M. Pulido-Mancera, T.~Zvolensky, M.~F. Imani, P.~T. Bowen, M.~Valayil, and
  D.~R. Smith, ``Discrete dipole approximation applied to highly directive
  slotted waveguide antennas,'' \emph{IEEE Antennas and Wireless Propagation
  Letters}, vol.~15, pp. 1823--1826, 2016.

\bibitem{lipworth2013JOSA}
G.~Lipworth, A.~Mrozack, J.~Hunt, D.~L. Marks, T.~Driscoll, D.~Brady, and D.~R.
  Smith, ``Metamaterial apertures for coherent computational imaging on the
  physical layer,'' \emph{J. Opt. Soc. Amer. A}, vol.~30, no.~8, pp.
  1603--1612, 2013.

\bibitem{li2010validityBorn}
J.~Li, X.~Wang, and T.~Wang, ``On the validity of born approximation,''
  \emph{Progress In Electromagnetics Research}, vol. 107, pp. 219--237, 2010.

\bibitem{lin1990invScatBorn}
F.~Lin and M.~A. Fiddy, ``Image estimation from scattered field data,''
  \emph{International Journal of Imaging Systems and Technology}, vol.~2,
  no.~2, pp. 76--95, 1990.

\bibitem{fear2002MWimaging}
E.~C. Fear, X.~Li, S.~C. Hagness, M.~Stuchly \emph{et~al.}, ``Confocal
  microwave imaging for breast cancer detection: Localization of tumors in
  three dimensions,'' \emph{Biomedical Engineering, IEEE Transactions on},
  vol.~49, no.~8, pp. 812--822, 2002.

\bibitem{liu2004nonlinearRecon}
Z.~Q. Zhang and Q.~H. Liu, ``Three-dimensional nonlinear image reconstruction
  for microwave biomedical imaging,'' \emph{IEEE Transactions on Biomedical
  Engineering}, vol.~51, no.~3, pp. 544--548, March 2004.

\bibitem{lipworth2015virtualizer}
G.~Lipworth, A.~Rose, O.~Yurduseven, V.~R. Gowda, M.~F. Imani, H.~Odabasi,
  P.~Trofatter, J.~Gollub, and D.~R. Smith, ``Comprehensive simulation platform
  for a metamaterial imaging system,'' \emph{Appl. Opt.}, vol.~54, no.~31, pp.
  9343--9353, Nov 2015.

\bibitem{papoulis2002probability}
A.~Papoulis and S.~U. Pillai, \emph{Probability, random variables, and
  stochastic processes}.\hskip 1em plus 0.5em minus 0.4em\relax Tata
  McGraw-Hill Education, 2002.

\bibitem{SARmanyVariables}
L.~M.~H. Ulander, H.~Hellsten, and G.~Stenstrom, ``Synthetic-aperture radar
  processing using fast factorized back-projection,'' \emph{IEEE Transactions
  on Aerospace and Electronic Systems}, vol.~39, no.~3, pp. 760--776, July
  2003.

\bibitem{yurduseven2016Aperiodic}
O.~Yurduseven, V.~R. Gowda, J.~N. Gollub, and D.~R. Smith, ``Printed aperiodic
  cavity for computational and microwave imaging,'' \emph{IEEE Microwave and
  Wireless Components Letters}, vol.~26, no.~5, pp. 367--369, May 2016.

\bibitem{yurduseven2015PSF}
O.~Yurduseven, M.~F. Imani, H.~Odabasi, J.~Gollub, G.~Lipworth, A.~Rose, and
  D.~R. Smith, ``Resolution of the frequency diverse metamaterial aperture
  imager,'' \emph{Progress In Electromagnetics Research}, vol. 150, pp.
  97--107, 2015.

\bibitem{scanSAR1981}
R.~K. Moore, J.~P. Claassen, and Y.~h.~Lin, ``Scanning spaceborne synthetic
  aperture radar with integrated radiometer,'' \emph{IEEE Transactions on
  Aerospace and Electronic Systems}, vol. AES-17, no.~3, pp. 410--421, May
  1981.

\bibitem{scanSAR1996}
A.~M. Guarnieri and C.~Prati, ``Scansar focusing and interferometry,''
  \emph{IEEE Transactions on Geoscience and Remote Sensing}, vol.~34, no.~4,
  pp. 1029--1038, Jul 1996.

\bibitem{nikolovaMicrowaveHolo}
M.~Ravan, R.~K. Amineh, and N.~K. Nikolova, ``Two-dimensional near-field
  microwave holography,'' \emph{Inverse Problems}, vol.~26, no.~5, p. 055011,
  2010.

\bibitem{neifeld2003featspecimag}
M.~A. Neifeld and P.~Shankar, ``Feature-specific imaging,'' \emph{Appl. Opt.},
  vol.~42, no.~17, pp. 3379--3389, 2003.

\bibitem{gehmFeatSpecific}
D.~V. Dinakarababu, D.~R. Golish, and M.~E. Gehm, ``Adaptive feature specific
  spectroscopy for rapid chemical identification,'' \emph{Opt. Express},
  vol.~19, no.~5, pp. 4595--4610, Feb 2011.

\end{thebibliography}

\end{document}